# Polarons in Lead-Halide Perovskites: What the Structural Evidence Does and Does Not Establish

Inhee Maeng[1,†], Young Mi Lee[2,†], Jinwoo Park[3], Seung Jae Oh[1], and Min-Cherl Jung[4,*]

[1] YUHS-KRIBB Medical Convergence Research Institute, College of Medicine, Yonsei University, Seoul 03722, Republic of Korea

[2] Beamline Department, Pohang Accelerator Laboratory, POSTECH, Pohang 37673, Republic of Korea

[3] Department of Physics, University of Seoul, Seoul 02504, Republic of Korea

[4] Global Institute of Future Technology, Shanghai Jiao Tong University, Shanghai 200240, China

[†]These authors contributed equally to this work.

[*]Corresponding author: Min-Cherl Jung, E-mail: mincherl.jung@sjtu.edu.cn

## ABSTRACT

Photogenerated carriers in lead-halide perovskites are widely believed to move as large polarons, whose accompanying lattice distortion is thought to screen the carrier from charged defects. Three recent findings call this picture into question. First, time-resolved structural probes now measure the photoinduced lattice distortion directly; yet calculations show that thermal motion of the lattice alone can raise the carrier effective mass—an increase previously taken as evidence for polarons—so that it remains unclear whether the lattice distortion around a carrier is really a polaron. Second, the carrier–lattice coupling strengthens as the material is thinned from 3D to 2D. In the two main families of 2D perovskites, Ruddlesden–Popper and Dion–Jacobson, the carrier couples to the lattice through different channels, and their contrasting behavior is set by which coupling channel dominates rather than by lattice stiffness alone. Third, the first direct test of defect tolerance does not point to polaron screening; its authors attribute defect tolerance to the soft, disordered lattice. New transport data show that polaron screening reduces scattering between carriers, but not that it protects carriers from defects. Polaron formation, even if established, therefore does not establish polarons as the cause of defect tolerance. Separating the two requires measuring the lattice distortion as a function of excitation density, over a range spanning the low densities of a solar cell under sunlight and the much higher densities used in most ultrafast experiments. On this basis, we propose experiments that can clarify how photogenerated carriers behave in lead-halide perovskites.

## I. INTRODUCTION

Lead-halide perovskites combine modest carrier mobilities with unusually long carrier lifetimes and diffusion lengths—properties realized in solution-processed films that, by the standards of conventional semiconductors, should contain many performance-limiting defects. Two types of explanation have been proposed for this paradox. The first is electronic: the valence-band maximum is an antibonding state formed by Pb 6*s*–halide *np* hybridization, which pushes the most common intrinsic defects toward shallow or resonant energy levels, so that structural imperfection does not necessarily create deep traps.[1,2] The second is dynamic: the Pb–X framework is soft and polarizable, and the A-site cations reorient on picosecond timescales. In such a lattice, a carrier couples to the longitudinal-optical (LO) phonon field and carries a nanometer-scale lattice distortion with it, and the resulting dynamic dielectric response screens the carrier from the Coulomb potential of charged point defects[3–5] (Fig. 1). Zhu and co-workers introduced this large-polaron picture, and Miyata and colleagues provided its first spectroscopic support.[3,4] For the past decade, this single quasiparticle has been invoked to explain slow carrier cooling, moderate mobility, radiative recombination, and the widely cited "defect tolerance."

Polarons in these materials have been reviewed from the viewpoint of device performance and the small-versus-large polaron classification,[6] and separately from the viewpoint of high-resolution characterization methods.[7] Our aim here is narrower: we ask which claims about polarons are supported by the evidence of the past five years. The most important change over this period is a higher standard of proof. Transient-absorption data alone cannot determine whether slow cooling originates from polaron formation, a hot-phonon bottleneck, or Auger reheating, because all three produce similar features in the same signal. Distinguishing these mechanisms requires three steps: (i) observing the lattice directly, (ii) tuning the electron–phonon coupling in a controlled way, and (iii) testing the defect-tolerance claim against a structural measurement.

Two recent results call for a fresh look at this picture. First, a direct experimental test of defect tolerance has been reported for the first time, and its authors do not adopt the polaron explanation.[8] Second, finite-temperature first-principles calculations now reproduce the effective-mass enhancement long read as evidence of polarons—without including a polaron at all.[9] The field therefore faces two related questions: does the polaron exist in the form usually pictured, and, if it does, do devices actually benefit from polaron screening? The alternative is that devices benefit from the softness and dynamic disorder of the lattice itself—the very properties that make polaron formation possible. A companion review from our group proposes that band-like transport, defect tolerance, and liquid-like lattice dynamics share a single origin in the soft, polar $[PbX_3]^-$ framework.[10] Here, we focus on what the structural evidence does and does not establish about the polaron itself.

We first define the range of materials considered. Most of the direct structural evidence in Sec. II comes from all-inorganic $CsPbBr_3$, in nanocrystal or single-crystal form.[9,11–14] By contrast, the screening picture of Fig. 1—and in particular its "ferroelectric large polaron" variant[5]—assigns a role to the reorienting organic A-site dipole. Yet no element-specific, time-resolved probe has tested, in any hybrid composition, whether the organic cation joins the polaronic distortion, merely follows it, or plays no role. We therefore use "lead-halide perovskites" for the material class and "hybrid" only where the composition requires it, and return to the A-site question in Secs. II D and VI.

This review is organized into three main sections (Secs. II–IV), followed by a synthesis in Sec. V and proposed experiments in Sec. VI. Section II surveys the time-resolved structural techniques that now resolve lattice distortion in real time and real space. Section III follows the crossover from large polarons to small and exciton polarons as the inorganic framework is thinned from 3D to 2D, and explains why Ruddlesden–Popper (RP) and Dion–Jacobson (DJ) perovskites differ for reasons beyond lattice rigidity. Section IV examines defect tolerance, for which direct evidence has now been reported—evidence that does not point to the polaron. Section V addresses the competing mechanisms and shows that a single variable, the carrier density, separates them. Section VI outlines the experiments—mostly structural, and mostly at fourth-generation X-ray sources—that would turn today's correlations into causal evidence.

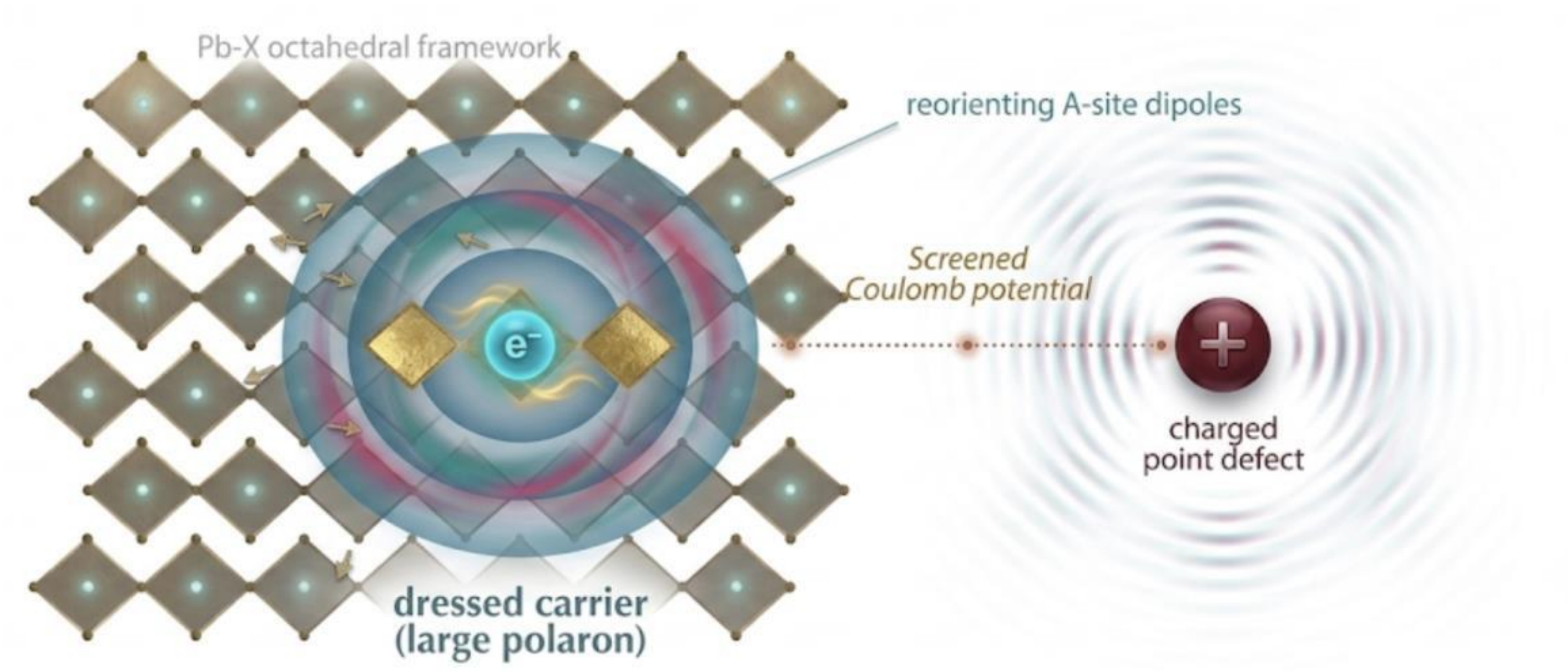

**FIG. 1.** Conceptual overview of the large-polaron screening picture. Left: corner-sharing Pb–X octahedral framework, drawn as a 2-dimensional array of [$PbX_6$] octahedra viewed along one crystallographic axis (gray squares; Pb at the center, halide at the corners). A photogenerated carrier ($e^-$) is surrounded by a nanometer-scale lattice distortion (large polaron, blue region); arrows indicate the accompanying reorientation of the A-site dipoles. Right: charged point defect, from whose Coulomb potential the lattice distortion and the reoriented dipoles screen the carrier. Two elements of this picture remain open: (i) whether the screening itself causes defect tolerance, or whether the lattice softness and dynamic disorder that permit the screening do (Secs. IV and V); and (ii) whether the organic A-site dipole takes part in the distortion at all. The most detailed structural evidence to date comes from all-inorganic $CsPbBr_3$, which has no organic dipole (Sec. II D).

## II. SEEING THE LATTICE MOVE: TIME-RESOLVED STRUCTURAL PROBES

The original evidence for large polarons was kinetic and optical.[3,4] Time-resolved optical Kerr-effect measurements on methylammonium lead bromide ($MAPbBr_3$) and $CsPbBr_3$ crystals revealed a nuclear response to carrier injection, with time constants of 0.29 and 0.7 ps, that was absent for sub-gap excitation.[4] Terahertz photoconductivity yielded effective masses and scattering parameters consistent with Fröhlich-dressed carriers,[15] and angle-resolved photoemission spectroscopy (ARPES) extended the argument to momentum space: the hole mass in $CsPbBr_3$ exceeds the calculated bare mass by roughly 50%.[13] Each of these experiments is sensitive, yet each sees the lattice distortion only indirectly, through its effect on the carriers. This indirect evidence is now in question, because finite-temperature calculations reproduce the same mass enhancement from thermal fluctuations of a strongly anharmonic lattice, without any polaron.[9]

Mobility has long faced the same problem: two different pictures of the carrier reproduce the same measured value. Fröhlich-coupling mobility calculations reproduce the measured room-temperature mobility of methylammonium lead iodide ($MAPbI_3$) without localization, whether the carrier is treated as a delocalized large polaron[16] or as a band state.[17] Consequently, mobility cannot distinguish a "large polaron" from a band carrier strongly scattered by LO phonons, and the effective mass suffers from the same ambiguity. Whatever the correct reading of the ARPES data, the lattice distortion should be measured directly rather than inferred from its effect on the carriers. Several probes now do so. Together, they span nearly four orders of magnitude in time resolution, and each reveals a different aspect of the lattice response (Fig. 2; Table I).

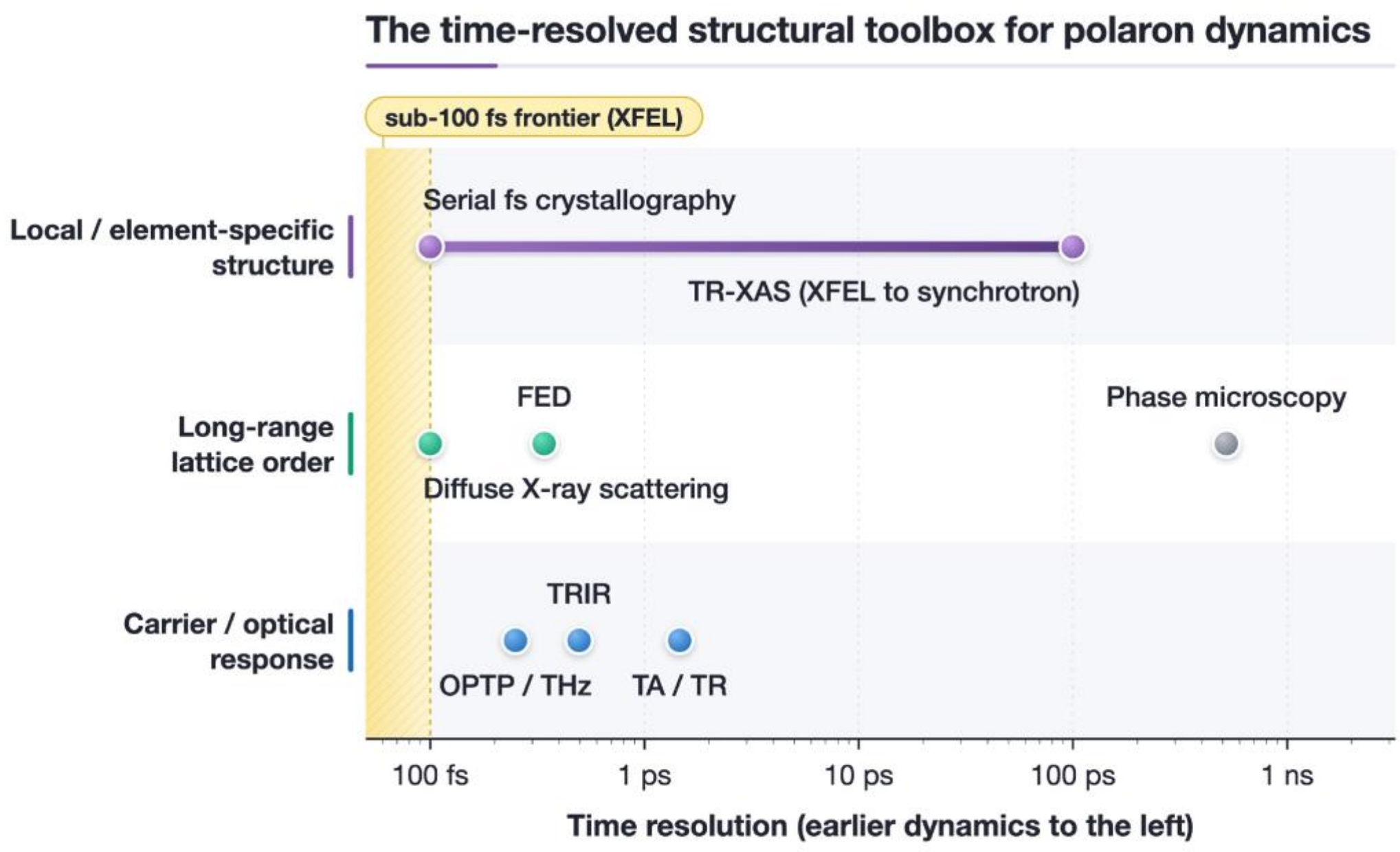

**FIG. 2.** Time-resolved structural toolbox, organized by time resolution (horizontal axis) and by the type of information returned (vertical axis). Optical probes—optical-pump–terahertz-probe spectroscopy (OPTP), transient absorption and reflection (TA/TR), and transient infrared spectroscopy (TRIR)—report on the carrier. Diffuse X-ray scattering and femtosecond electron diffraction (FED) report on long-range lattice order, whereas serial femtosecond crystallography and time-resolved X-ray absorption spectroscopy (TR-XAS) are local and element-specific. Phase microscopy extends the map to the nanosecond range. The sub-100 fs window (shaded), accessible at X-ray free-electron lasers (XFELs), remains the main experimental frontier.

### *A. Diffuse X-ray Scattering and the Polaronic Strain Field*

Femtosecond diffuse X-ray scattering provided the first momentum-resolved view of the distortion. Guzelturk and co-workers imaged nanometer-scale strain fields that form, expand radially, and relax around photoexcited carriers in $MAPbBr_3$ single crystals, extracting quantitative distortion amplitudes and correlating them with transient changes in the carrier effective mass.[18] This experiment delivered the first direct measurement of the size, shape, and relaxation time of the distortion attributed to the large polaron. Notably, the strain field builds up over tens of picoseconds—much more slowly than the sub-picosecond dressing seen optically. The optical and structural signals therefore report different processes and should not be treated as the same observable (Fig. 3). Complementary measurements resolve the build-up of the polaron in α-$FAPbI_3$ (FA, formamidinium) in the time domain,[19] and momentum-resolved theory now predicts what a forming polaron should look like in a diffraction pattern,[20] so that these predictions can be tested directly in experiments rather than compared only qualitatively.

“Figure adpted from [18], reproduced with permission in the final published version”

**FIG. 3.** Femtosecond diffuse X-ray scattering of the polaronic strain field: the nanometer-scale strain field forms and relaxes around photoexcited carriers, and its distortion amplitude correlates with transient changes in the carrier effective mass. Reproduced with permission from Ref. 18. Copyright 2021, Springer Nature.

### *B. Ultrafast Electron Diffraction*

Femtosecond electron diffraction (FED) is directly sensitive to both coherent and incoherent lattice motion. In $MAPbI_3$ thin films, it captured the development of rotational disorder in the halide-octahedral sublattice: the Bragg intensities decayed with a time constant of about 10 ps and the pair-distribution function changed with a time constant of about 27 ps, at a carrier density of $2.3\times10^{19}$ $cm^{-3}$.[21] The method has since been applied to $CsPbBr_3$ nanocrystals, where it tracked sub-picosecond lattice dynamics and revealed a light-induced distortion that appears on a fluence-dependent timescale of 0.38–1.2 ps, but no signature of a multi-picosecond hot-phonon bottleneck[12] (Fig. 4). Optical data cannot separate the mechanisms in this way. This negative result, however, was obtained under one set of conditions and should not be generalized: a hot-phonon bottleneck has been reported in rigid 2D DJ films at high excitation density.[22] Reconciling the two results requires samples of matched composition, dimensionality, and carrier density (Sec. V).

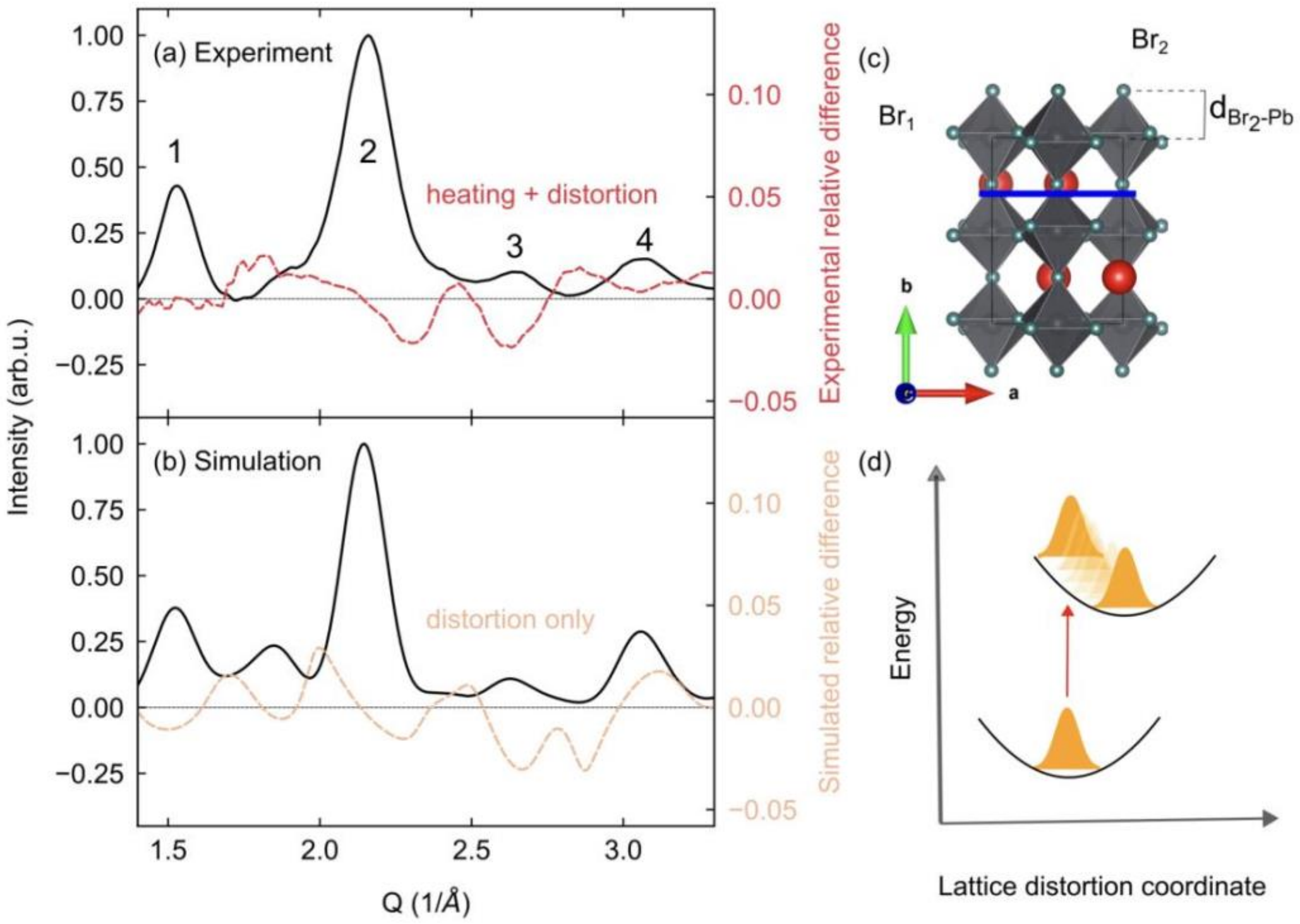

**FIG. 4.** Femtosecond electron diffraction of $CsPbBr_3$ nanocrystals: a sub-picosecond, fluence-dependent lattice distortion accompanies exciton-polaron formation, and no multi-picosecond hot-phonon bottleneck is detected under these conditions. Reproduced with permission from Ref. 12. Copyright 2023, American Chemical Society.

### *C. Element-Specific X-ray Absorption and Serial Crystallography*

Whereas diffraction measures long-range lattice order, time-resolved X-ray absorption spectroscopy (TR-XAS) measures the local structure around one selected element. Working at the Br *K*- and Pb $L_3$-edges, Cannelli and co-workers quantified the photoinduced polaronic distortion in inorganic lead-halide nanocrystals by combining temperature-dependent kinetics with ab initio simulation of the core-hole final state. Their analysis shows which atoms move and by how much relative to one another: the axial Br displacement is about six times the equatorial Br displacement and about 2.5 times the Pb off-center displacement.[11] Element selectivity is decisive here because it can, in principle, identify the sublattice that hosts the carrier and test whether the A-site cation takes part. The screening hypothesis requires precisely this site-resolved information, which optical spectroscopy cannot provide.

Serial femtosecond crystallography has since reached the single-particle limit, resolving the deformation field of a single electron–hole pair in individual $CsPbBr_3$ quantum dots.[14] Two of its findings are directly relevant here. First, the distortion is best described as a mixed object—a delocalized electron paired with a localized hole—rather than as a large polaron. Second, resonant single-exciton excitation produces a qualitatively different deformation from non-resonant multi-carrier excitation in the same material. The latter finding bears on how the field should collect its data, and we return to it in Sec. V.

All-optical coherent methods are beginning to close the gap between optical and structural timescales. Multidimensional coherent spectroscopy on $CsPbBr_3$ quantum dots, analyzed with a lineshape theory that relates the anti-diagonal linewidth to a collective lattice-polarization order parameter, yields an exponential polaron build-up time of 150 fs,[23] while two-pulse photon-echo measurements on $CsPbI_3$ nanocrystals resolve a fully coherent exciton–polaron ladder with dephasing times of several hundred picoseconds at cryogenic temperature.[24] Neither is a structural measurement, yet both provide sub-picosecond timing benchmarks against which the first X-ray free-electron laser (XFEL) data below 100 fs can be judged.

The momentum-space debate has also moved on. Temperature-dependent ARPES on $CsPbBr_3$ now resolves orthorhombic replica bands and splittings of several hundred meV that are tied to octahedral tilts; octahedral tilts alone therefore narrow the bands, and this effect must be separated from any polaronic mass enhancement.[25] Time-resolved ARPES on a quasi-2D hybrid perovskite finds no small-polaron self-trapping and no photoinduced Rashba splitting up to 120 ps.[26] Moreover, a special-displacement framework that treats polymorphism, anharmonicity, and electron–phonon coupling on an equal footing yields temperature-dependent effective masses without an explicit polaron.[27] Together with the finite-temperature calculations of Ref. 9, these results do not disprove the polaron, but they remove the effective mass from the list of observables that can prove it. Raman spectroscopy carries a related caveat: the broad central peak long linked to local polar fluctuations[28] has been re-attributed to disorder-induced second-order acoustic scattering,[29] and should not be cited as a polaron-specific signature without independent confirmation.

*TABLE I.* What each time-resolved structural probe can and cannot establish about the polaron.

| **Probe** | **Observable** | **Time resolution** | **Materials studied to date** | **Establishes** | **Cannot establish** |
|---|---|---|---|---|---|
| fs diffuse X-ray scattering (XFEL) | Momentum-resolved diffuse intensity → real-space strain field | fs (XFEL) | $MAPbBr_3$ single crystal[18] | Size, shape and relaxation time of the strain field; correlation with effective mass | Element identity of the displaced atoms; A-site participation |
| fs electron diffraction (FED/UED) | Bragg and diffuse intensities; coherent and incoherent lattice motion | ~100–300 fs | $MAPbI_3$ films[21]; $CsPbBr_3$ nanocrystals[12] | Sub-ps onset of distortion; presence/absence of a multi-ps non-thermal phonon population | Local bond-length changes; distinction from thermal disorder without a dark baseline |

| Probe | Observable | Time resolution | Materials studied to date | Establishes | Cannot establish |
|---|---|---|---|---|---|
| TR-XAS (Br *K*, Pb $L_3$) | Element-specific local structure and oxidation state | ~100 fs (XFEL) to ~100 ps (synchrotron) | $CsPbBr_3$ nanocrystals[11] | Which sublattice hosts the distortion; relative displacement of each atom type, with ab initio core-hole modeling | Long-range order; organic-cation response (requires soft-X-ray N/C *K*-edge) |
| Serial fs crystallography (single particle) | Full 3D diffraction of individual nanocrystals | ~100 fs | $CsPbBr_3$ quantum dots[14] | Deformation field of one exciton; resonant vs. non-resonant excitation differ qualitatively | Bulk and thin-film behavior; hybrid compositions |
| THz emission / OPTP | Carrier response and low-frequency lattice resonances | ~100 fs | α-$FAPbI_3$ films[19]; $MAPbI_3$ (Refs. 15 and 30) | Polaron build-up in the time domain; density regimes of trap, bimolecular and Auger recombination | Structural distortion amplitude (inferred, not measured) |
| Multidimensional coherent / photon-echo spectroscopy | Lineshape and coherence of the exciton–polaron | ~10 fs | $CsPbBr_3$,[23] $CsPbI_3$ (Ref. 24) nanocrystals | Build-up time (~150 fs) and coherence of the dressed state | Any structural quantity directly |

### *D. The A-Site Question: What Hybrid Compositions Still Lack*

TR-XAS in $CsPbBr_3$ is consistent with the hole residing on the halide sublattice, but it probes only the Br *K*- and Pb $L_3$-edges, so that the A-site cation ($Cs^+$) is not probed directly; no such measurement exists for $MA^+$ or $FA^+$. The A-site matters because the two variants of the large-polaron hypothesis differ on exactly this point: the Fröhlich picture requires only the polar $[PbX_3]^-$ framework, whereas the ferroelectric-large-polaron picture[5] requires the organic dipoles to order transiently around the carrier. Ultrafast electron diffraction on $MAPbI_3$ (Ref. 21) and diffuse scattering on $MAPbBr_3$ (Ref. 18) show that the inorganic framework responds, and THz emission from α-$FAPbI_3$ resolves a low-frequency resonance associated with the $FA^+$ sublattice during polaron build-up.[19] None of these measurements, however, is element-specific for nitrogen or carbon. Time-resolved soft-X-ray absorption at the N *K*-edge is technically demanding but within reach of current XFEL and fourth-generation synchrotron end-stations. Such a measurement would establish whether the organic cation participates or is a bystander, and we list it among the priorities in Sec. VI.

Taken together, these probes agree that a real lattice distortion of measurable amplitude forms on timescales from sub-picoseconds to tens of picoseconds, while differing on details that depend on composition, dimensionality, and excitation density. Two frontiers remain. The first is temporal: existing studies resolve polaron formation over picoseconds to tens of picoseconds but rarely reach the first 100 fs, the range of the optically detected build-up,[23] for which XFELs and fourth-generation synchrotrons are the natural tools (Sec. VI). The second is interpretive: a photoinduced distortion is not by itself a polaron, because a soft anharmonic lattice fluctuates strongly even in the dark.[9,27] Distinguishing a polaron from this thermal background requires the excitation-density dependence discussed in Sec. V.

## III. THE DIMENSIONAL CROSSOVER: FROM LARGE TO SMALL AND EXCITON POLARONS

If the 3D bulk lies in the weak-to-intermediate Fröhlich regime, thinning the inorganic framework is the cleanest way to tune the electron–phonon coupling. Confining carriers to a slab a few octahedra thick between bulky organic spacers increases both quantum and dielectric confinement, which strengthens the coupling and drives the system toward small- and exciton-polaron behavior. The crossover is qualitative as well as quantitative: the character of the quasiparticle itself changes (Fig. 5). First-principles calculations that go beyond small periodic supercells now show that halide perovskites can host small polarons, large polarons, and charge-density-wave-like states within a single framework,[31] providing the experimental crossover with a theoretical vocabulary that it previously lacked. A macroscopic theory of polarons and exciton polarons in two-dimensional polar media gives a further reason why the crossover is qualitative rather than gradual: in 2D, both the LO-phonon dispersion and the Lyddane–Sachs–Teller relation take a different form, so that the Fröhlich coupling itself acquires features with no 3D counterpart.[32]

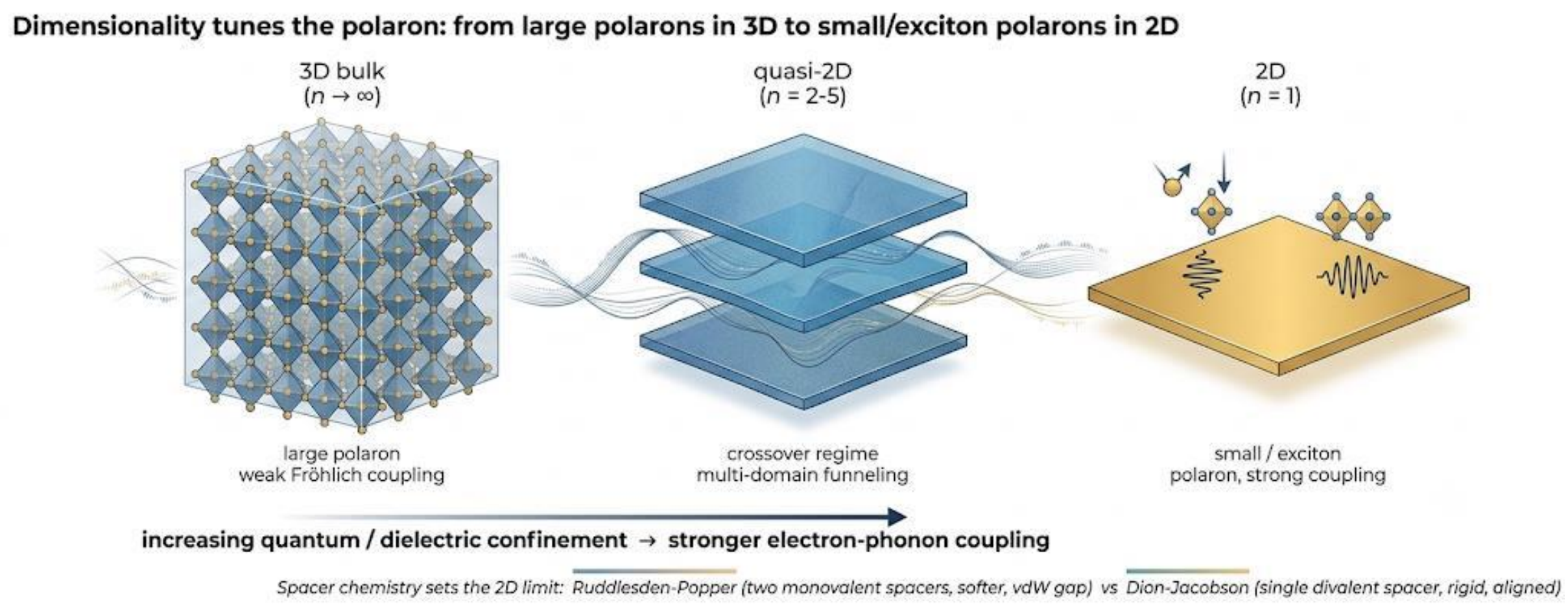

**FIG. 5.** Dimensional crossover. As the number of inorganic layers $n$ decreases from the 3D bulk through quasi-2D to the 2D ($n = 1$) limit, quantum and dielectric confinement strengthen the electron–phonon coupling, moving the system from weakly coupled large polarons to strongly coupled small and exciton polarons. The spacer chemistry (Ruddlesden–Popper versus Dion–Jacobson) sets the balance between polar (Fröhlich) and deformation-potential coupling, and hence where a given compound lands in the 2D limit.

### *A. Why RP and DJ Perovskites Behave Differently, and Why Rigidity Is Not the Reason*

The two dominant 2D families differ structurally, and this difference determines the coupling mechanism. RP perovskites employ two monovalent spacer cations held together by van der Waals contact, giving a softer, offset, loosely bound stack. DJ perovskites employ a single divalent spacer that bridges adjacent inorganic slabs directly, giving a more rigid and better-aligned lattice.

Rigidity is commonly assumed to settle the matter: a stiff lattice should suppress the local relaxation that dresses a carrier. DJ compounds, however, do not follow this rule. Rigid, harmonic-lattice DJ compounds show a persistent exciton that is only weakly dressed,[33] whereas DJ compounds with giant deformation potentials localize carriers into genuine small polarons about two unit cells across, with transport that differs markedly from the 3D case[34] (Fig. 6). The contradiction disappears once the two coupling channels are considered separately. Fröhlich coupling to polar LO modes scales with the dielectric response of the lattice and is indeed suppressed by rigidity. Deformation-potential coupling to acoustic modes scales instead with the band-edge shift per unit strain; in the $n = 1$ member of one DJ series it exceeds $10^2$ eV—about 2 to 6 times the values for $n = 2$ and 3, and some thirty times that of conventional 2D or 3D perovskites.[34] Rigidity therefore predicts weak dressing only when the polar channel dominates. What places a compound on the crossover is which coupling channel dominates, not which structural family it belongs to.

Nonadiabatic molecular dynamics on DJ compounds makes the same point from the theoretical side: rigid, high-dipole spacers do not simply switch the coupling off, but shift the dominant coupling from high- to low-frequency modes, thereby prolonging hot-carrier lifetimes.[35] Phonons shape the dynamics of DJ films in other ways as well. Exciton polarons form and are strongly dressed in some DJ films,[22] and biexcitonic Auger recombination occurs on sub-picosecond to few-picosecond timescales, at a rate set by exciton–phonon coupling and dielectric confinement.[36] DJ films therefore span the range from weakly to strongly dressed excitons.

The softer RP lattice, by contrast, favors stronger and faster exciton-polaron dressing through the polar channel,[37,38] and the dressed state remains dynamic rather than permanently self-trapped.[39] The acoustic channel is not absent in RP perovskites either: microscopic theory and magneto-optics on $(PEA)_2PbI_4$ (PEA, phenethylammonium) identify acoustic-phonon-driven scattering from bright to dark excitons as the relaxation bottleneck on the ~20 ps scale.[40] The same tightly bound divalent-spacer architecture that slows halide-ion migration in DJ films relative to RP films[41,42] thus also governs their polar electron–phonon coupling—a case in which a processing advantage and a fundamental coupling parameter share a single structural origin.

Recent transport data support what the optical data imply. In RP germanium-halide single-crystal transistors, $(PEA)_2GeI_4$ exhibits thermally activated small-polaron hopping, whereas the isostructural $(PEA)_2SnI_4$ exhibits band-like transport; cation engineering that weakens the electron–phonon coupling restores band-like transport with an order-of-magnitude gain in mobility.[43] Because the layer architecture is the same in both compounds, it is the coupling strength, rather than the rigidity, that determines the transport regime. Self-trapped excitons in $(PEA)_2PbI_4$ also reach room-temperature diffusion lengths exceeding those of free excitons, sustained by long lifetimes and photon recycling.[44] The small-polaron label thus describes the local distortion, not the macroscopic transport.

### *B. What the Recent Spectroscopy Shows*

Studies from 2024–2026 have rapidly filled in this crossover. Vibronic progressions in the absorption lineshape of 2D perovskites have been interpreted directly as polaron signatures[45]; exciton-polaron formation and its effect on hot-carrier relaxation have been resolved in rigid DJ films[22]; time-resolved photoluminescence (TRPL) has quantified large exciton-polaron shifts in RP layers[37]; anisotropic electron–phonon coupling has been mapped in 2D films[46]; and the bright-exciton fine structure has been tracked continuously from the 2D to the bulk limit.[47] Exciton–phonon coupling strengthens systematically with the inorganic-layer number $n$ in mixed-phase quasi-2D samples,[48] and modeling has begun to explain the full 2D-to-3D electron–phonon trend and its consequences for hot-carrier cooling.[49] Ultrafast studies of RP films further link spacer chemistry and synthetic route to exciton-polaron stability and population balance.[50]

The most direct experimental map of the $n$-dependence to date comes from transient mid-infrared spectroscopy on the PEA-based series. At $n = 1$, an intense distortion produces a featureless continuum, indicating rapid self-trapping; at $n = 2$, resolved Franck–Condon progressions of ~10 meV modes appear. The coupling mechanism itself thus changes with $n$, in addition to becoming weaker.[51] Impulsive stimulated Raman scattering on $(BA)_2PbI_4$ (BA, butylammonium) adds a collective effect: coherent excitation of a Higgs amplitude mode drives the layer transiently toward a higher-symmetry tetragonal phase. Above-gap pumping produces a four-fold smaller spectral shift than below-gap pumping, which the authors attribute to photoexcited carriers exerting a counteracting displacive force on the octahedra.[52]

Two cautions apply to this rapid progress. First, vibronic sidebands are the most frequently cited polaron fingerprint in 2D spectra, yet photonic effects—namely an excitonic stop band and thin-film interference—can produce reflectance plateaus and thickness-dependent fringes in RP perovskites. These effects must be ruled out before a spectral feature is assigned to phonons or polarons, although a single stop band does not by itself explain the higher-energy sidebands, whose origin remains open.[53] Second, the coherent-phonon and Raman observables discussed above report on the phonon subsystem, not directly on the dressing of the carrier. As in 3D, the decisive experiment is a time-resolved structural measurement, and it has not yet been performed on an $n$-series.

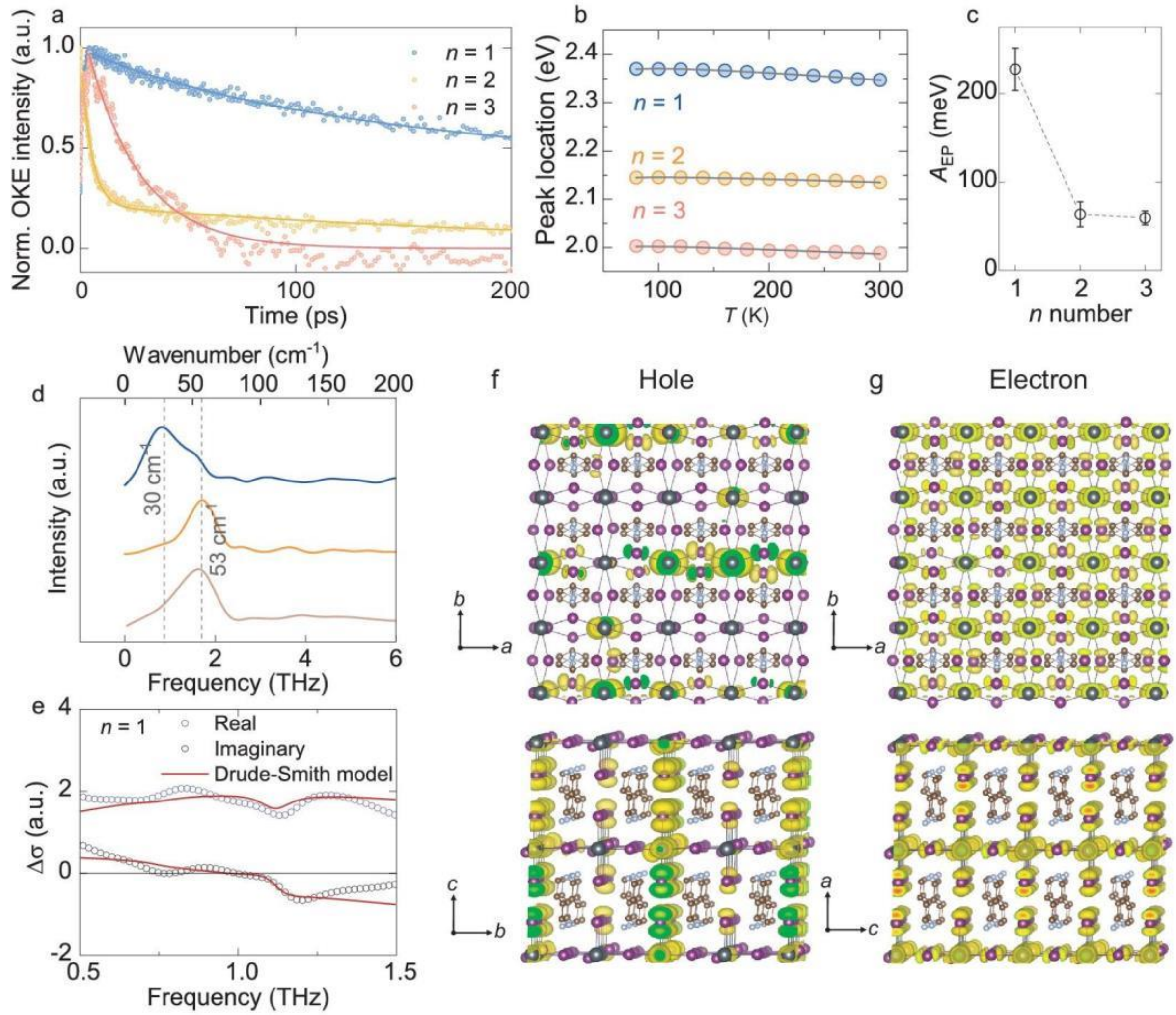

**FIG. 6.** Spectroscopic evidence for small-polaron formation driven by a giant deformation potential in a Dion–Jacobson 2D lead-halide perovskite: strong dressing can occur in a rigid lattice when the non-polar coupling channel dominates. Reproduced from Ref. 34 under the Creative Commons CC BY license.

Table II summarizes the crossover as a working map. The boundaries are deliberately soft: quasi-2D films are multi-domain by construction, so that a single sample can host several regimes at once, and energy and charge funneling between domains then becomes part of the physics rather than an artifact.

*TABLE II.* Dimensionality-dependent polaron regimes across the 3D-to-2D lead-halide perovskite series.

| **Structural class** | **Representative system** | **Dominant polaron regime** | **Diagnostic signature** | **Key refs** |
|---|---|---|---|---|
| 3D bulk ($n \to \infty$) | $MAPbI_3$, $MAPbBr_3$, $FAPbI_3$, $CsPbBr_3$ | Large polaron; weak-to-intermediate Fröhlich coupling; long-range Coulomb screening | Long-lived THz/TA response; fs diffuse X-ray strain fields; ~150 fs optical build-up | 3,4,15,18,19,23 |
| Quasi-2D ($n$ = 2–5) | $(PEA)_2(MA)_{n-1}Pb_nI_{3n+1}$ multi-domain films | Crossover regime; domain-dependent | TRPL/TA tracking low-$n$ → | 47–49,51 |

| Structural class | Representative system | Dominant polaron regime | Diagnostic signature | Key refs |
|---|---|---|---|---|
| | | cooling and inter-domain funneling; coupling mechanism changes with $n$ | high-$n$ transfer; $n$-dependent e–ph coupling; resolved Franck–Condon progressions at $n = 2$ | |
| 2D ($n = 1$), Ruddlesden–Popper | Monoammonium-spaced $PbI_4$ layers | Strongly dressed (exciton-)polaron via the polar channel; softer, offset stacking favors local relaxation; acoustic channel governs bright–dark scattering | Phonon coherences; large exciton-polaron shift in TRPL; TRIR continuum at $n = 1$; coherent Higgs-mode response | 37–40,45,50–52 |
| 2D ($n = 1$), Dion–Jacobson | Rigid diammonium-spaced $PbI_4$ layers | Persistent weakly dressed exciton when the polar channel dominates; small polaron when deformation-potential coupling dominates | Vibronic progression in absorption; short, localized small-polaron lifetime under large deformation potential; fast biexcitonic Auger | 22,33–36 |

## IV. POLARON SCREENING AND THE DEFECT-TOLERANCE QUESTION

For years, defect tolerance was treated almost as an axiom: structural defects were assumed not to become electronic traps, with large-polaron screening as the most popular explanation. This assumption rested on computation rather than measurement, and the computational case is itself disputed. Zhang and co-workers showed that in $CsPbI_3$ the energetically favorable $Pb_I$ and $I_{Pb}$ antisites lie deep in the gap, so that the common statement that all energetically favorable defects are shallow is incorrect; what renders these antisites harmless is strong anharmonicity, which makes them inefficient nonradiative centers.[54] Later work by the same group concludes that halide perovskites are not defect tolerant, because other defects, such as the iodine interstitial, are strong nonradiative centers.[55]

Direct experimental evidence has been reported only recently. Jasti and co-workers performed what is, to our knowledge, the first direct test, comparing the structural quality of 2D, 2D–3D, and 3D Pb-iodide single crystals with high-sensitivity optoelectronic measurements that probe the bulk rather than the surface.[8] For each composition, they grew higher- and lower-quality crystals, quantifying structural quality by rocking-curve X-ray diffraction and optoelectronic quality by photoluminescence together with modulated sub-bandgap surface photovoltage. The two quantities are uncorrelated: the rocking-curve widths of the higher- and lower-quality crystals differ by factors from roughly four to more than one hundred, yet the integrated sub-bandgap photovoltage responses are similar, differing by only about three times the experimental error. This lack of correlation constitutes the experimental evidence for defect tolerance in these crystals, and it holds across the whole series—2D, 2D–3D, and 3D alike (Fig.

7).

Three caveats accompany this result. First, it was obtained on macroscopic single crystals and therefore says nothing about surfaces and grain boundaries, which are precisely where polycrystalline devices fail. Second, the transient-photoluminescence data could not be fitted by either mono- or bi-exponential functions, and the extracted lifetimes run contrary to simple expectation: the higher-quality $MAPbI_3$ crystal decays faster than the lower-quality one. The authors report this openly and build no conclusions on it. Third, rocking curves probe extended (periodic) structural defects, so that the measurement does not cover thermodynamically determined point defects.

This result, however, does not prove that polaron screening causes defect tolerance. Jasti and co-workers did not adopt the large-polaron explanation; instead, they point to the low deformation energy of the soft Pb–halide lattice and its dynamic local disorder,[56,57] either of which can mask the electronic consequences of structural defects without invoking a dressed carrier. The causal link to the polaron thus remains inferred, not measured. Parallel work describes defect tolerance as a dynamic phenomenon under working conditions, driven by the soft, anharmonic lattice as much as by any single quasiparticle,[58–60] with self-healing of the lattice as a further, distinct contribution.[61] Computations have meanwhile examined the surface, to which the single-crystal test is blind: machine-learning force-field simulations of $CsPbBr_3$ surfaces at 300 K find that roughly 70% of the surface-state energies remain within 0.2 eV of the valence-band maximum, even at cleaved bonds, because atomic motion prevents deep states from remaining isolated.[62]

The dynamic-disorder argument, however, now requires an important qualification. For the chlorine vacancy in $CsPbCl_3$, multi-task force-field dynamics show that the optical transition level oscillates strongly on picosecond timescales, yet neither the nonradiative capture barriers nor the thermodynamic transition levels are affected, and the defect obeys equilibrium thermodynamics.[63] Fluctuation of the defect level is therefore not by itself a tolerance mechanism. Whether fluctuation protects or exposes a defect depends on the capture pathway, which must consequently be computed defect by defect.[54,55]

A second distinction is equally important and has often been blurred. New density-resolved terahertz and microwave conductivity measurements spanning $10^{13}$–$10^{20}$ $cm^{-3}$ show that large-polaron screening delays the mobility roll-off caused by carrier–carrier scattering to ~$10^{19}$ $cm^{-3}$ (Ref. 64)—a density at which the carrier lifetime has already collapsed through multi-carrier recombination. This is direct evidence that polaron screening reduces Coulomb scattering between carriers. It is not evidence that polaron screening suppresses nonradiative capture at defects, which is what defect tolerance concerns, and the two effects should not be used interchangeably.

The practical conclusion is that the field now possesses both ingredients needed to settle the question: (i) a structural-quality axis that can be varied deliberately—composition, dimensionality, controlled non-stoichiometry, or irradiation dose—and (ii) a set of probes that measure the polaron amplitude directly. What is missing is an experiment that varies the structural quality and measures the polaron amplitude in the same sample.

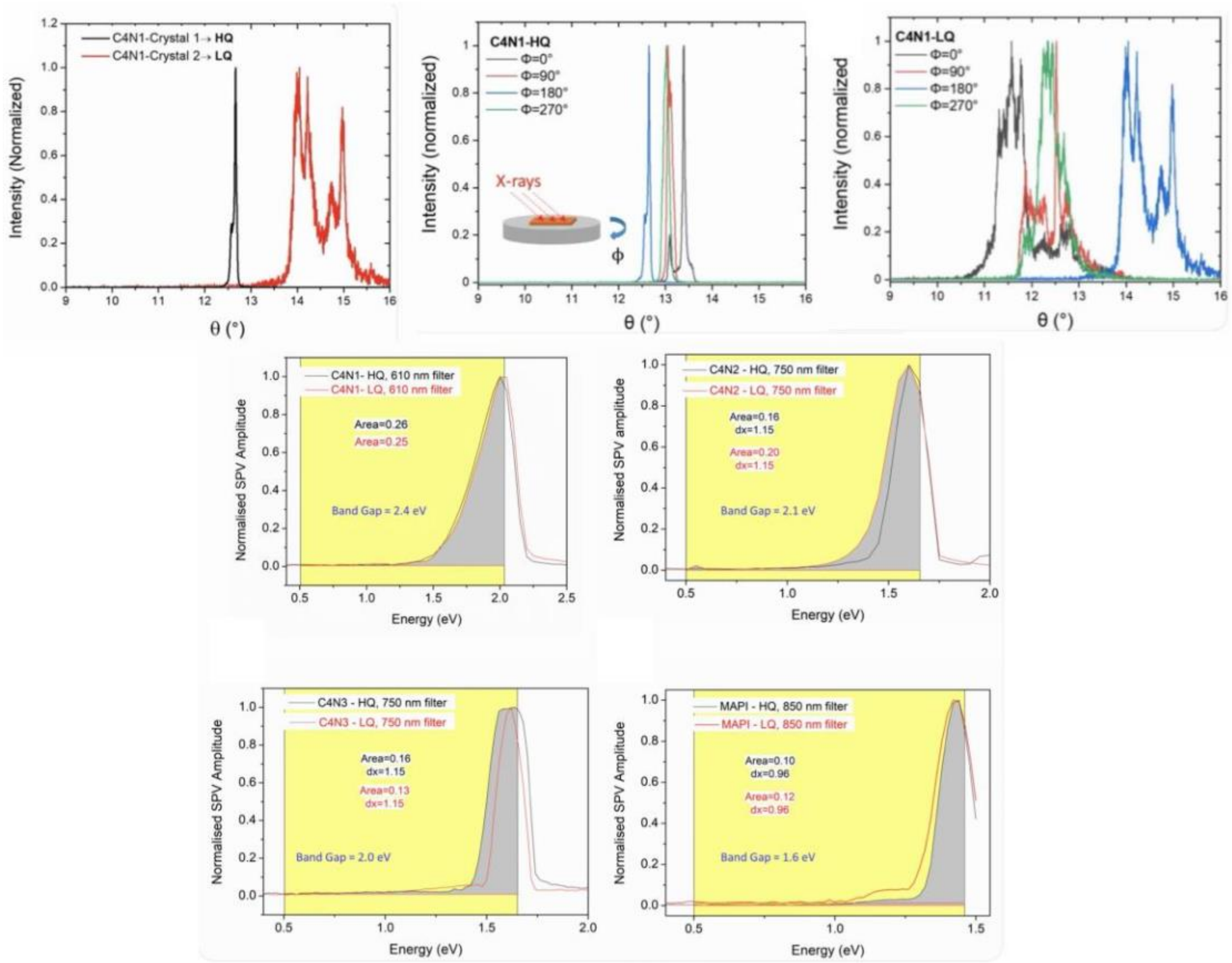

**FIG. 7.** First direct experimental test of defect tolerance. Rocking-curve X-ray diffraction (structural quality) and modulated sub-bandgap surface photovoltage (optoelectronic quality) for higher- and lower-quality crystals of 2D, 2D–3D, and 3D Pb-iodide perovskites: rocking-curve widths differing by a factor of four to more than one hundred produce no correspondingly large difference in the sub-bandgap defect response. Reproduced from Ref. 8 under the Creative Commons CC BY-NC-ND license.

## V. DISENTANGLING COMPETING MECHANISMS

Any account of slow carrier cooling must consider three candidate mechanisms that produce similar optical signatures: large-polaron screening, the hot-phonon bottleneck, and Auger reheating. The earliest transient-absorption studies of $MAPbI_3$ identified the hot-phonon bottleneck at carrier densities above ~$5\times10^{17}$–$10^{18}$ $cm^{-3}$,[65,66] and Auger heating was shown to take over at around $10^{19}$ $cm^{-3}$.[67] The polaron interpretation offered a different explanation,[3] and was later linked to low-fluence cooling, with the crossover to high-fluence bottleneck cooling attributed to a Mott transition at which the polarons overlap ($n \geq 10^{18}$ $cm^{-3}$).[68] Consequently, at the high excitation densities historically needed for adequate signal, the same slow-cooling signature has been assigned to different mechanisms. A recent survey of optical hot-carrier probes warns that variations in experimental implementation and data analysis can lead to conflicting or even erroneous interpretations.[69]

Two approaches can now distinguish these mechanisms. The first is structural fingerprinting, which exploits the qualitatively different predictions that the three mechanisms make for the lattice. Polaron formation produces a specific distortion, observed as strain-field diffuse scattering[18]; in $CsPbBr_3$ nanocrystals, the electron-diffraction response, in which only some Bragg peaks shift, differs from simple heating, which would shift all peaks and reduce all intensities.[12] A phonon bottleneck would appear as a long-lived non-thermal phonon population, which time-resolved diffraction can detect and did not find in $CsPbBr_3$ nanocrystals.[12] Auger reheating heats the electronic system rather than the lattice directly,[67] and must therefore be separated from the other two by its density dependence, not by a lattice fingerprint. Phonon-resolved work on a related anharmonic semiconductor, $Bi_2O_2Se$, identifies a specific anharmonic phonon mode that mediates polaron formation, moving the discussion from "coupling to LO phonons" to named modes.[70]

These fingerprints, however, depend on the experimental conditions. The same DJ films in which exciton-polaron formation is clearly resolved also develop a hot-phonon bottleneck once the excitation density is raised,[22] and RP films show a bottleneck that strengthens or weakens with the synthetic route at fixed composition.[50] Polaron formation and a phonon bottleneck are thus not mutually exclusive alternatives but processes with different density thresholds, and no single experiment at a single fluence can rank them. Structural fingerprinting identifies which mechanism operates under specified conditions; it does not establish which mechanism is "correct" in general.

The second approach is the more important one: to measure outside the high-fluence regime in which most measurements have been made. The data are hardest to interpret at carrier densities far above those reached under one-sun operation, which correspond to $10^{15}$–$10^{16}$ $cm^{-3}$.[71,72] Electron diffraction on $CsPbBr_3$ nanocrystals used 0.7–5.6×$10^{19}$ $cm^{-3}$,[12] and the classic hot-carrier cooling study on $MAPbI_3$ used densities from 1.6×$10^{17}$ up to 6.4×$10^{18}$ $cm^{-3}$.[65] The density landscape is now mapped well enough to locate the boundaries (Fig. 8). Temperature- and density-dependent optical-pump–terahertz-probe measurements on $MAPbI_3$, spanning $10^{14}$–$10^{19}$ $cm^{-3}$ and 78–315 K, separate trap-, bimolecular-, and Mott-dominated regimes, and locate the onset of ultrafast polaron–polaron annihilation, likely Auger-type, above ~$10^{18}$ $cm^{-3}$, where the polaron wavefunctions begin to overlap.[30] Complementary conductivity data place the polaron-screened mobility roll-off near $10^{19}$ $cm^{-3}$.[64] The onset of polaron overlap and the mobility roll-off are distinct observables with different thresholds, and Fig. 8 marks them separately. Below ~5×$10^{17}$ $cm^{-3}$, no hot-phonon bottleneck is observed[65,66]; polaron–polaron annihilation[30] is reported only above ~$10^{18}$ $cm^{-3}$, and Auger heating[67] at ~$10^{19}$ $cm^{-3}$. In the range below these thresholds, the polaron, if it forms, is the only excitation-dependent mechanism considered here that remains. Fluence- and temperature-dependent measurements covering the one-sun-equivalent regime, combined with a quantitative readout of the lattice strain, would therefore allow each mechanism to be identified under device-relevant conditions.

Excitation density has already been shown to change the result: single-exciton serial crystallography finds a deformation field qualitatively different from that produced by non-resonant multi-carrier excitation in the same material.[14] A higher excitation density does more than scale the signal—it changes what is measured.

Studies that question the polaron picture reach the same conclusion from the opposite starting point. Ensemble Monte Carlo simulations reproduce the ultra-long cooling times of tin-based perovskites from band filling and a hot-phonon bottleneck alone, with no polaron term,[73] and a critical perspective finds that many hot-carrier claims—including fast hot-carrier extraction, quasi-ballistic transport, and the bottleneck itself—either fail replication or have simpler explanations.[74] None of these studies rules the polaron out; each shows that, at the densities where the polaron has usually been invoked, it was not needed. Both routes described above are within reach of existing XFEL and synchrotron end-stations. We therefore regard the density-resolved structural experiment as the most feasible high-impact problem in this area.

Section IV raised the same problem for structural data, where several mechanisms explain one null result. The problem is more severe there because the list of candidates is longer. At least four mechanisms compete to explain defect tolerance: (i) benign defect energetics arising from the antibonding valence-band maximum[1,2,54]; (ii) polaron screening[3,4]; (iii) the low phonon energies of the soft Pb–halide lattice, which make nonradiative capture a slow multiphonon process[75]; and (iv) dynamic local disorder, which shifts in-gap levels by up to an electron volt on picosecond timescales.[56,63] These mechanisms are not independent: a soft, anharmonic lattice drives dynamic disorder[76] and, in the polaron picture, is also what permits a large polaron.[10] Demonstrating that polarons form, however clearly, therefore does not demonstrate that they cause defect tolerance.

One variable separates these mechanisms. Benign defect energetics is a property of the ground-state band structure; lattice softness and static disorder are present in the dark; and thermal fluctuation persists without photoexcitation and can mimic polaronic mass enhancement.[9] Only the polaron requires excitation. Experiments that vary the carrier density while measuring the lattice strain therefore separate the cooling mechanisms and the defect-tolerance mechanisms simultaneously—something no other single variable achieves (Fig. 9).

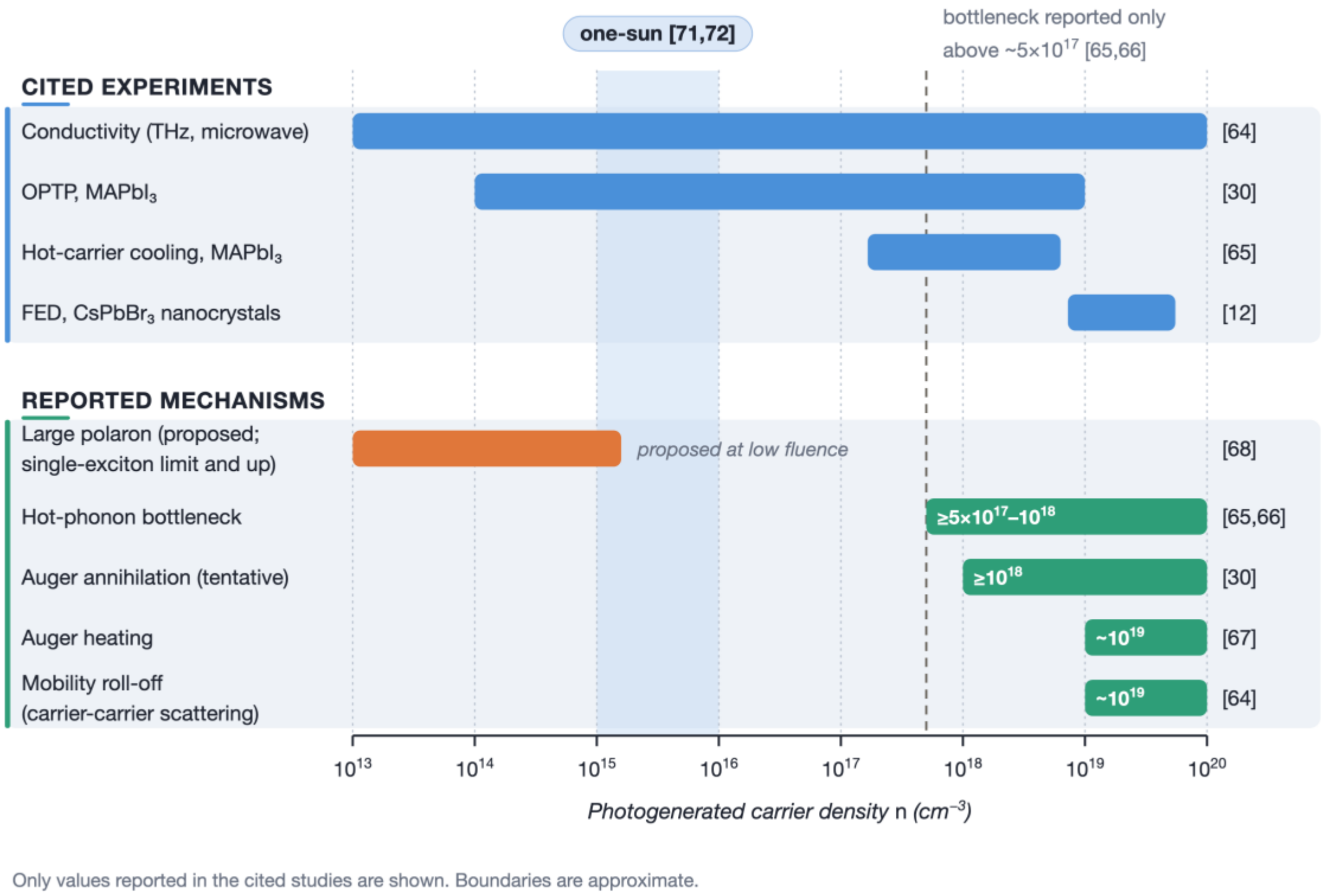

**FIG. 8.** Carrier-density ranges over which the competing mechanisms compared in Fig. 9 have been reported. One-sun operation of a 3D absorber corresponds to $10^{15}$–$10^{16}$ cm$^{-3}$.[71,72] Experimental ranges: electron diffraction on $CsPbBr_3$ nanocrystals, 0.7–5.6×$10^{19}$ cm$^{-3}$ (Ref. 12); optical-pump–terahertz-probe measurements on $MAPbI_3$, $10^{14}$–$10^{19}$ cm$^{-3}$ (Ref. 30); conductivity measurements, $10^{13}$–$10^{20}$ cm$^{-3}$ (Ref. 64); and the hot-carrier cooling study on $MAPbI_3$, 1.6×$10^{17}$ to 6.4×$10^{18}$ cm$^{-3}$.[65] Reported thresholds: hot-phonon bottleneck above ~5×$10^{17}$–$10^{18}$ cm$^{-3}$ (Refs. 65 and 66); Auger annihilation above ~$10^{18}$ cm$^{-3}$ (tentative assignment)[30]; Auger heating at ~$10^{19}$ cm$^{-3}$ (Ref. 67); and mobility roll-off from carrier–carrier scattering near $10^{19}$ cm$^{-3}$.[64] Lattice softness, dynamic disorder, and benign defect energetics are ground-state properties present in the dark. Polaron formation is drawn as a proposed mechanism at low fluence,[68] down to the single-exciton limit.[14] The boundaries are approximate and depend on composition, temperature, and morphology.

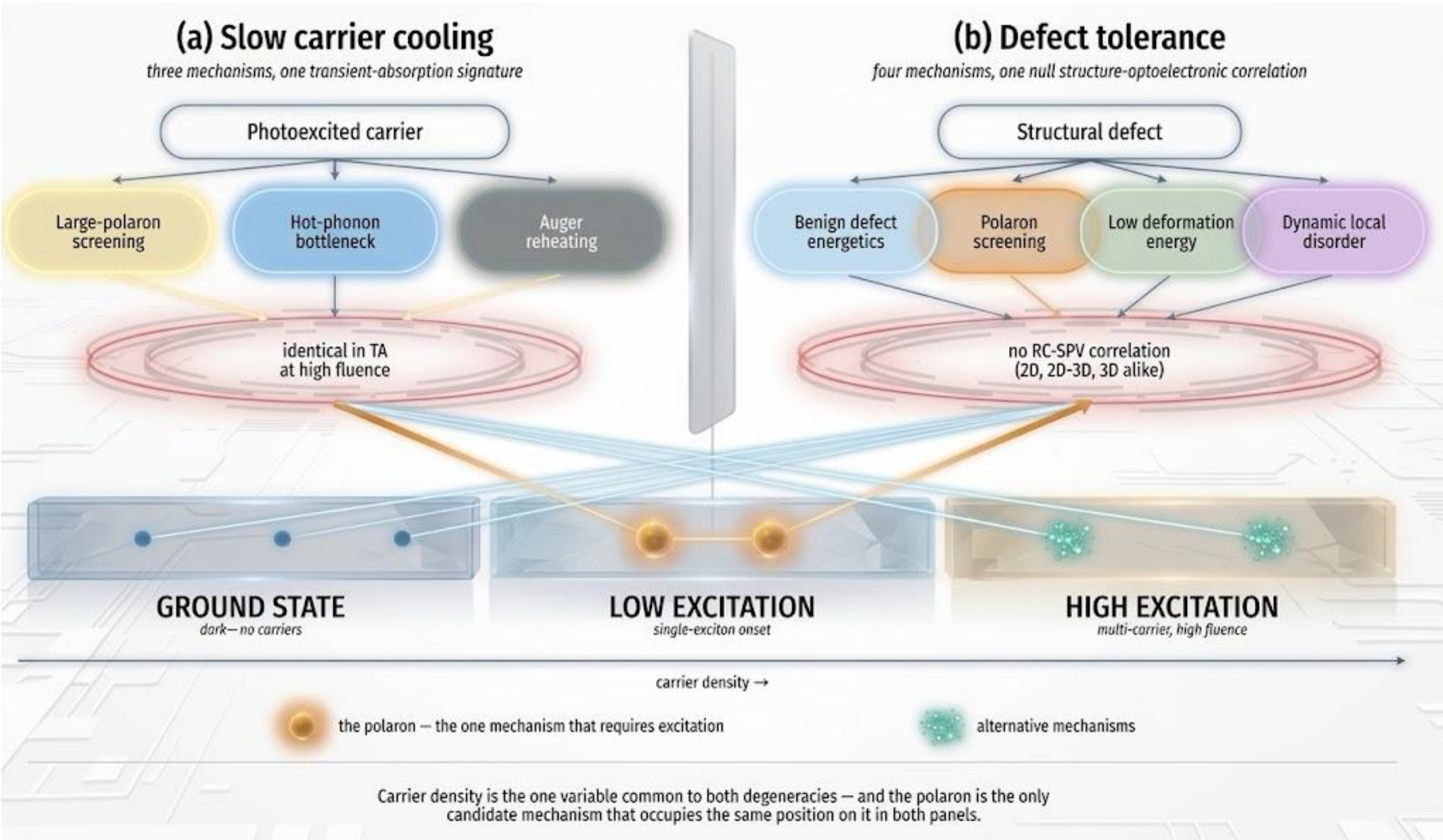

**FIG. 9.** Two sets of competing explanations and the variable that separates them. (a) Three mechanisms produce a similar transient-absorption signature of slow carrier cooling. (b) Four mechanisms produce the same null result for defect tolerance, namely no correlation between structural and optoelectronic quality. Both sets are placed on a single shared axis, the carrier density, running from the unexcited ground state through the single-exciton limit to the high-fluence, multi-carrier regime historically used to obtain adequate signal. The proposed polaron (gold in both panels) is the only mechanism occupying the same low-density position in both panels; every alternative (teal) is either a ground-state lattice or defect property or requires a substantially higher excitation density. Varying the carrier density while measuring the lattice strain is therefore the single experiment that addresses both panels at once (Sec. V).

## VI. OUTLOOK AND CONCLUSION

The three threads of this review—direct structural evidence, the dimensional crossover, and defect tolerance—converge on a coherent experimental program rather than a list of unrelated gaps, with the excitation-density axis of Sec. V as its common thread. We highlight six priorities and one conceptual caveat.

**Sub-100 fs nucleation.** Femtosecond diffuse X-ray scattering and diffraction should be pushed below 100 fs at an XFEL or a fourth-generation synchrotron, with the goal of capturing the earliest coherent lattice response in benchmark 3D compositions. FED has so far detected the light-induced distortion in $CsPbBr_3$ nanocrystals on a 0.38–1.2 ps timescale.[12] Figure 10 places the relevant processes on a common time axis; the ~150 fs build-up time now extracted optically for $CsPbBr_3$ quantum dots[23] sets the target that the structural measurement must resolve.

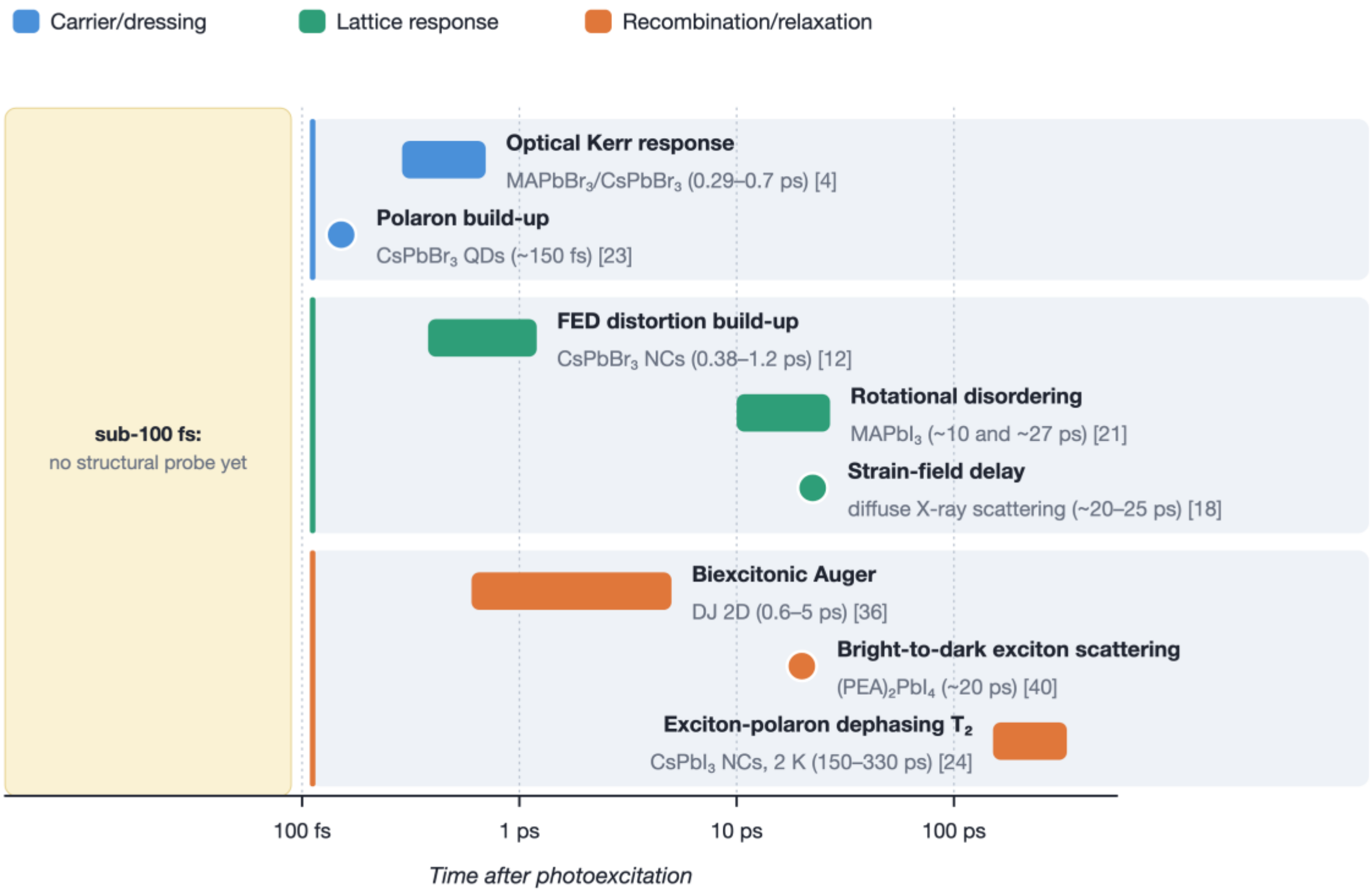

**FIG. 10.** Timescales of the processes discussed in this review, compiled from the cited experiments. Optical Kerr response to carrier injection: time constants of 0.29 ps ($MAPbBr_3$) and 0.7 ps ($CsPbBr_3$).[4] Optically detected polaron build-up: ~150 fs (marker).[23] Light-induced lattice distortion detected by FED in $CsPbBr_3$ nanocrystals: build-up time constants of 0.38–1.2 ps, depending on fluence.[12] In $MAPbI_3$, the Bragg intensities decay with a time constant of about 10 ps and the pair-distribution function changes with a time constant of about 27 ps.[21] Polaronic strain field imaged by diffuse X-ray scattering: delayed by about 20–25 ps at small wavevector.[18] Biexcitonic Auger recombination in DJ 2D perovskites: 0.6–5 ps.[36] Bright-to-dark exciton scattering in $(PEA)_2PbI_4$: ~20 ps (marker).[40] Exciton–polaron dephasing in $CsPbI_3$ nanocrystals at 2 K: $T_2$ = 150–330 ps.[24] The shaded region marks the sub-100 fs window that no structural probe has yet reached. All values are those reported in the cited studies; conditions such as composition, temperature, and excitation density differ between studies.

**A ground-state baseline.** Every claim of a photoinduced distortion requires a dark control of comparable precision, because thermal fluctuations in these lattices are large enough to reproduce polaronic observables on their own.[9,27] Temperature-dependent measurements of the unexcited lattice, made with the same instrument and the same analysis, should therefore accompany every pump–probe dataset. This is the least expensive priority on this list.

**Polaron amplitude versus quantified defect density, across the density gap.** The distortion amplitude should be measured directly in single crystals with systematically varied and quantified non-stoichiometry or irradiation dose, and then compared with defect-limited recombination—from the one-sun-equivalent regime upward, and not only at the fluences where Auger and bottleneck processes are already active (Fig. 8). Benign defect energetics, lattice softness, and dynamic disorder are ground-state properties, whereas the polaron exists only under excitation; an excitation-dependent structural readout is therefore what separates them. As Sec. IV shows, the field still lacks this experiment.

**A matched RP/DJ homologous series.** The crossover should be mapped across $n = 1–5$ at fixed fluence and temperature, with RP and DJ spacers compared directly. The variable to report is the ratio of Fröhlich to deformation-potential coupling rather than the rigidity, since it is this ratio that places a compound on the crossover map. Measured in this way, the crossover becomes a phase diagram rather than a collection of case studies.

**The organic cation, element by element.** Time-resolved soft-X-ray absorption at the N *K*-edge (and, where feasible, the C *K*-edge) should be performed on MA- and FA-based crystals. For hybrid compositions, this would provide the same site-resolved information that TR-XAS already delivers for the halide and lead sublattices of $CsPbBr_3$,[11] and would decide between the Fröhlich and ferroelectric variants of the large-polaron hypothesis[4,5] on structural rather than dielectric grounds.

**From single crystals to manufacturable films.** These measurements should be extended to device-relevant interfaces and to scalable polycrystalline films, in which grain boundaries, interfacial fields, and processing strain modify both the polaron and its screening role—and about which the only direct defect-tolerance test to date says nothing.

**A conceptual caveat.** The word "polaron" itself requires care. Recent theory shows that in strongly anharmonic halide lattices, one-electron–two-phonon (nonlinear Fröhlich) terms contribute significantly to the electron mobility of $CsPbI_3$, reducing it by about 10% at room temperature. Treatments that assume linear electron–phonon coupling may therefore be incomplete, even where a polaron forms.[77]

Polaron research in lead-halide perovskites has matured from indirect optical interpretation into a structural science, and the structural data complicate the simplest picture. Time-resolved structural probes now measure the photoinduced lattice distortion directly, and the dimensional crossover from 3D to 2D shows that the character of the carrier–lattice coupling depends on which coupling channel—polar Fröhlich or deformation-potential—dominates, rather than on lattice rigidity alone. The first direct evidence for defect tolerance does not identify the polaron as its cause; the soft, dynamically disordered lattice that makes polarons possible may, on its own, be a sufficient explanation. Finite-temperature theory, moreover, now reproduces some of the polaron's signature observables without the quasiparticle. To our knowledge, the only process in which polaron screening has been shown experimentally to matter is carrier–carrier scattering,[64] and defect tolerance does not operate through that process. None of this shows that polarons are absent. It shows that some reports have interpreted one set of observables as though only one mechanism could produce them, at densities where several do. Separating the mechanisms is now the central question, and existing XFEL and fourth-generation synchrotron end-stations can address it.

## ACKNOWLEDGMENTS

## Declaration of Generative AI Use

During the preparation of this work, the authors used generative AI tools (Anthropic Claude and Google Gemini) to assist in drafting the layout of the original conceptual schematics shown in Figs. 1, 2, 5, 8, 9, and 10, and in English proofreading of the text. All scientific content, labeling, and structural representations in these figures were specified, verified, and edited by the authors, who take full responsibility for their accuracy.

## Note on Figure Reproduction

*Figures 1, 2, 5, 8, 9, and 10 are original schematics prepared for this work. Figures 3, 4, 6, and 7 are reproduced from prior publications. Ref. 34 (Natl. Sci. Rev.) is Creative Commons CC BY and may be reproduced freely with citation. Ref. 8 (PNAS) is CC BY-NC-ND and may be reproduced only as published, without modification. Refs. 18 (Springer Nature) and 12 (ACS) are subscription articles, for which reprint permission for scholarly reuse is obtained through RightsLink (Copyright Clearance Center), normally at no charge. Licensing has been reviewed separately by the corresponding authors.*

## AUTHOR DECLARATIONS

### *Conflict of Interest*

The authors have no conflicts to disclose.

### *Author Contributions*

I.M. and Y.M.L. contributed equally to this work. M.-C.J. conceived and supervised the review. I.M. and Y.M.L. surveyed the literature and drafted the manuscript; I.M. and S.J.O. prepared the schematic figures. Y.M.L. and J.P. verified the references. All authors discussed, reviewed and approved the final manuscript.